\documentclass[sigchi]{acmart}

\usepackage{booktabs}
\usepackage{commath}
\usepackage{threeparttable}

\newcommand\symUserSet{U}
\newcommand\symItemSet{I}

\newcommand\symUserInteractionSet{S}

\DeclareMathOperator{\sign}{sign}
\renewcommand\vec{\mathbf}

\begin{document}

\title{Binary Latent Representations for Efficient Ranking}
\subtitle{Empirical Assessment}
\author{Maciej Kula}
\email{maciej.kula@gmail.com}
\date{\today}
\settopmatter{printacmref=false, printccs=false, printfolios=true}
\setcopyright{none}
\acmConference[]{}{}{}



\begin{abstract}
Large-scale recommender systems often face severe latency and storage constraints at prediction time. These are particularly acute when the number of items that could be recommended is large, and calculating predictions for the full set is computationally intensive. In an attempt to relax these constraints, we train recommendation models that use binary rather than real-valued user and item representations, and show that while they are substantially faster to evaluate, the gains in speed come at a large cost in accuracy. In our Movielens 1M experiments, we show that reducing the latent dimensionality of traditional models offers a more attractive accuracy/speed trade-off than using binary representations.
\end{abstract}

\maketitle

\section{Introduction}
Industry ranking and recommendation systems need to scale to tens or hundreds of millions of items and users. As the number of items available to be recommended grows large, efficiently ranking the item catalogue to produce top-ranked recommendations becomes increasingly challenging. With large catalogues, ranking latency (in on-line settings) and compute and storage costs (in pre-compute settings) place significant constraints on the design of the entire recommender system.

In an on-line setting, where recommendation latency directly impacts the user experience, system designers have to work within a strict latency budget. Respecting that budget often means trading off model accuracy for speed, either via smaller and less expressive models (with lower-dimensionality representations) or through aggressive use of heuristics to exclude large classes of candidate items from scoring. For example, Pinterest uses a combination of heuristics and candidate-generation models to select approximately 1000 pins, from billions of available pins, as input to its related pins ranking system \citep{liu2017related}. Similarly, approximate nearest neighbour search techniques are used at YouTube \citep{covington2016deep} to reduce the number of candidate videos that need to be scored for each user.

The challenge of low-latency scoring leads many systems to rely on a pre-compute system, where recommendations are calculated in advance and stored for serving at a later time. The primary disadvantage of such a system is its inability to react dynamically to user actions (such as new interactions that allow the recommendations to be refined) and recommendation context (location, time of day or year). This leads to often complex architectures that mix fairly simple on-line models, computationally expensive offline models, and intermediate complexity near-line algorithms where model updating is important \citep{amatriain2013big}.

Additionally, while a pre-compute solution removes model prediction from the critical path, it still requires substantial investment in computational and storage resources to calculate and cache the predictions of the model. Often the amount of computation required makes timely updating of the cached results impossible. At Pinterest, pre-calculation of related pins results was so demanding that ranking of different segments of the catalogue had to be staggered in time, leading to stale results and reduced ease of development \citep[Section 5.5]{liu2017related}. \citet{koenigstein2012efficient} estimate that computing recommendations for the Yahoo! Music dataset \citep{dror2011yahoo} with a 50-dimensional latent factor model woud take over 135 hours.

To alleviate these constraints in both online and offline settings, we evaluate the use of binary item and user representations in learning-to-rank matrix factorization models. Following the approach of \citet{rastegari2016xnor}, we estimate binary user and item representations that are several times faster to score and requiring a fraction of the memory (for the same latent dimensionality). Unfortunately, they are substantially less accurate than standard learning-to-rank approaches. The resulting speed-accuracy trade-offs favour simply reducing the latent dimensionality of traditional models when evaluation speed is desired.

We speculate that binary representations are more useful in case of more complex models (such as deep convolutional models) where a single prediction requires many more floating operations than relatively compact bilinear recommender models.

\section{Binary latent representations}
\label{sec:approach}
\citet{rastegari2016xnor} introduce XNOR-Networks, an approach that combines 1-bit quantization of fully-connected and convolutional layers with representing dot products through scaled XNOR and bitcounting operations.

We adapt their approach to the recommendation task by learning binary versions of user and item vectors within a learning-to-rank factorization model. This allows us to use binary instead of floating point operations for computing recommendation rankings, which has significant speed and memory use advantages. In a real-valued $n$-dimensional factorization model, item and user representations take $4n$ bytes of memory, and computing a single prediction takes $2n$ floating point operations. In an equivalent binary model, $\frac{3}{32}n$ binary operations are required for each prediction (XOR, negation, and bitcounting), and the representations take $\frac{1}{8}n$ bytes of memory. Naive calculations suggest that, for the same latent dimensionality, a binary model would be over 20 times faster to score, and take less then 5\% of memory to store its representations.

Naturally, using binary instead of real-valued representations entails a loss of representational power, and the resulting models can be expected to be less accurate. Nevertheless, binary models may still be very useful in a production system, for two reasons. Firstly, they may offer the system designer a more favourable trade-off between speed and accuracy than simply changing the number of latent dimensions in a real-valued model. If latency is the binding constraint, switching to a binary model may offer better or comparable speed with better ranking performance. If accuracy is the priority, a binary model may offer equivalent ranking performance while being faster to score. Secondly, many production systems resort to heavy use of heuristics when selecting candidates to be scored by their ranking systems (as scoring all candidates would be infeasible). To the extent that these heuristics are suboptimal, replacing them with a binary representation model would improve the overall accuracy of the system.

To quantify the trade-offs involved, we construct and fit both a real-valued and a binary version of simple implicit learning-to-rank factorization model. We introduce the notation and the models below.

Let $\symUserSet$ be the set of users, and $\symItemSet$ be the set of items. Each user interacts with a number of items $\symUserInteractionSet^+$; the set of all remaining items is denoted by $\symUserInteractionSet^-$.

In a traditional real-valued latent factor model, the prediction $r_{ui}$ for any user-item interaction pair $(u, i) \in \symUserSet \times \symItemSet$ is given by
\begin{equation}
r_{ui} = \vec{u}_u \cdot \vec{i}_i + b_u + b_i
\end{equation}
where $\vec{u}_u$ denotes the user and $\vec{i}_i$ the item $n$-dimensional latent vectors, and $b_u$ and $b_i$ the user and item biases.

\citet{rastegari2016xnor} show that the dot product of two real-valued vectors can be approximated in the binary domain by (1) binarizing the input vectors using the sign function (so that the results lie in $\{1, -1\}$), (2) taking their dot product, and (3) scaling the result by the average magnitude of the vectors' elements. Letting
\begin{equation}
\alpha = \frac{1}{n}\norm{\vec{i}_{i}}_{\ell1}
\end{equation}
and
\begin{equation}
\beta = \frac{1}{n}\norm{\vec{u}_{u}}_{\ell1}
\end{equation}
represent the scaling factors, the approximation is given by
\newcommand\binaryApproximation{ \alpha \beta\left(\sign(\vec{u}_u) \cdot \sign(\vec{i}_i)\right)}
\begin{equation}
\vec{u}_u \cdot \vec{i}_i \approx \binaryApproximation
\end{equation}
Applying this to the latent-factor model, the prediction for a user-item pair is given by
\begin{equation}
r_{ui} = \binaryApproximation + b_u + b_i
\end{equation}
Note that the scaling factors $\alpha$ and $\beta$ need to be stored (in addition to the latent representations and biases) to compute predictions in the binary model. Their use also implies a constant cost of two floating-point multiplications per dot product when using binary representations.

The model is trained using backpropagation, and so we use a smooth approximation to the derivative of the sign function to facilitate training. Following \citet{courbariaux2016binarized}, we define
\begin{equation}
  \frac{d\sign}{dw}=
  \begin{cases}
    1 & \text{if}\ \abs{w} \leq 1 \\
    0 & \text{otherwise}.
  \end{cases}
\end{equation}

We use two loss functions to train the model:
\begin{itemize}
\item Bayesian personalised ranking (BPR, \citet{rendle2009bpr}), and
\item adaptive sampling maximum margin loss, following \citet{weston2011wsabie}.
\end{itemize}
For both loss functions, for any known positive user-item interaction pair $(u, i)$, we uniformly sample an implicit negative item $j \in \symUserInteractionSet^-$. For BPR, the loss for any such triplet is given by
\begin{equation}
1 - \sigma\left(r_{ui} - r_{uj}\right),
\end{equation}
where $\sigma$ denotes the sigmoid function.
The adaptive sampling loss is given by
\begin{equation}
\abs{1 - r_{ui} + r_{uj}}_{+}.
\end{equation}
For any $(u, i)$ pair, if the sampled negative item $j$ results in a zero loss (that is, the desired pairwise ordering is not violated), a new negative item is sampled, up to a total of $k$ attempts. This leads the model to perform more gradient updates in areas where its ranking performance is poorest.

For the purposes of this paper, we focus on a simple bilinear collaborative filtering model, and do not use any external metadata information or model the sequential nature of the data. However, the binary dot product approach generalizes beyond simple factorization models, and can be applied as a component of any model whose final scoring step involves a dot product between user and item representations. In particular, models using recurrent \citep{hidasi2015session} or convolutional \citep{lynch2015images} item or user representations can easily be augmented to use binary dot products in the final ranking stages.

\section{Experiments}
\begin{table*}[htbp]
\begin{threeparttable}
\caption{Movielens 1M results}
\label{tb:results}
\centering
\begin{tabular}{rrrrrrrr}
\toprule
Dimension &   MRR & Binary MRR & MRR ratio\tnote{1} &   PPMS\tnote{2} & Binary PPMS & PPMS ratio\tnote{3} & Memory use ratio\tnote{4} \\
\midrule
       32 & 0.077 &      0.059 &     0.768 & 87,758 &     264,146 &      3.010 &            0.091 \\
       64 & 0.075 &      0.066 &     0.878 & 45,992 &     220,833 &      4.802 &            0.062 \\
      128 & 0.076 &      0.071 &     0.935 & 23,870 &     166,252 &      6.965 &            0.047 \\
      256 & 0.078 &      0.071 &     0.908 & 12,284 &     190,131 &     15.477 &            0.039 \\
      512 & 0.080 &      0.073 &     0.912 &   6074 &     122,637 &     20.190 &            0.035 \\
     1024 & 0.080 &      0.075 &     0.936 &   3056 &      61,301 &     20.056 &            0.033 \\
\bottomrule
\end{tabular}
\begin{tablenotes}
\small{
\item[1] Ratio of binary model MRR to real-valued model MRR
\item[2] Predictions Per Millisecond: how many items can be scored per millisecond
\item[3] Ratio of binary PPMS to real-valued PPMS
\item[4] Ratio of memory required to store binary vs. real-valued parameters
}
\end{tablenotes}
\end{threeparttable}
\end{table*}
To assess the accuracy-speed trade-offs enabled by the XNOR approach, we conduct an experiment on the Movielens 1M dataset \citep{harper2016movielens}. The dataset contains 1 million ratings from 6000 users on 4000 movies. Because the computational speed improvements offered by the binary representations are linear in the size of the catalogue, the results on this dataset should be indicative of results that can be obtained on larger datasets. At the same time, the dataset's small size enables us to easily test hyperparameter configurations and estimate models with high latent dimensions.

\subsection{Experimental setup}
We randomly divide the dataset into a training, test, and validation set. In order to build a picture of the accuracy-speed trade-offs afforded by real-valued and binary latent models, we explicitly build models for between 32 and 1024 latent dimensions. For every latent dimensionality, we conduct a random search over the hyperparameter space, and pick optimal initial learning rate, loss function, L2 penalty, minibatch size and number of training epochs for each algorithm based on their performance on the test set. The final results are obtained from ranking interactions from the validation set. We use mean reciprocal rank (MRR) as a measure of ranking quality.

Benchmark results are measured on an Intel Xeon E5-2686v4 CPU by running a 500 repetitions of scoring 100,000 items and averaging the results.

\subsection{Implementation}
The model is implemented in PyTorch and trained using the nVidia K40 GPUs. The implementation is only marginally more complex than a standard learning-to-rank model, and is accomplished by simply replacing the user-item vector dot product operation with its binary counterpart. The change results in a small increase in training time.

During training, the embedding parameters are stored and updated as single precision floating point values. Similarly, the XNOR dot product is carried out using floats in $\{1, -1\}$. The initial learning rate, loss function, L2 penalty, minibatch size and number of training epochs are treated as model hyperparameters. All models are trained using the Adam training rate schedule \citep{kingma2014adam}.

The prediction code runs on the CPU and is implemented in C using Intel X86 AVX2 SIMD intrinsics. SIMD (Single Instruction Multiple Data) instructions allow the CPU to operate on multiple pieces of data in parallel, achieving significant speedups over the scalar version. We use explicit intrinsics rather than compiler autovectorization to ensure that neither the real-valued nor the binary prediction code is unfairly disadvantaged by the quality of compiler autovectorization. 

The real-valued prediction code is implemented using 8-float wide fused multiply-add instructions (\texttt{\_mm256\_fmadd\_ps}). In the binary version, the real-valued embedding parameters used in training are discarded, and the derived 1-bit weights are packed into 32-bit integer buffers. The XNOR dot product is implemented using 8-integer wide XOR operations (256 binary weights are processed at a time), followed by a popcount instruction to count the number of on bits in the result. We use the \texttt{libpopcnt} \citep{mula2016faster} library for the bit counting operations.

Both versions use 32-bit aligned input data to utilise aligned SIMD register load instructions. The code is compiled using GCC 4.8.4 for the AVX2-enabled Broadwell architecture.

The source code for both model training and prediction is available on Github at \url{https://github.com/maciejkula/binge}.

\subsection{Results}
\label{sec:results}
Table \ref{tb:results} summarises the results of the Movielens 1M experiment. Each row presents the best MRR (mean reciprocal rank) results for both real-valued and binary models of a given dimensionality. MRR ratio denotes the fraction of the real-valued MRR the binary model achieves for that dimensionality; PPMS (predictions per millisecond) ratio denotes how many more predictions the binary model can compute per millisecond.

Results on scoring speed and memory use broadly confirm the naive calculations from section \ref{sec:approach}, converging to over 20 times faster scoring at around 3\% of memory as latent dimensionality increases. Memory savings plateau as latent dimensionality increases and storing the binary scaling factors becomes less important.

As expected, for the same dimensionality a binary model achieves lower ranking accuracy. On average, the accuracy loss when moving from a continuous to a binary model of the same latent dimensionality is around 11\%, varying between 6\% and 23\%.

Unfortunately, while continuous models retain good accuracy as latent dimensionality decreases, binary models' representational power sharply deteriorates. Moving from the 1024 to 32 dimensions in the continuous model implies a 29 times increase in prediction speed at the expense of a modest 4\% decrease in accuracy. This compares favourably to switching to binary representations: moving to a 1024-binary dimensional representation implies a sharper accuracy drop at 6\% in exchange for a smaller 20 times increase in prediction speed. Moving to 32 dimensions yields a further speed gains at 86 times, but at the cost of a considerable loss of accuracy at 26\%.

The results suggest that latent recommender models have good representational power even at low latent dimensionalities. This advantage is lost when using binary embeddings, which need to be high-dimensional to achieve comparable accuracy. At the same time, binary representations' speed advantage is only evident at high dimensions: at low dimensions, only floating-point operations can use enjoy the advantages of SIMD operations, and the fixed overhead of binary scaling factors $\alpha$ and $\beta$ constitutes a larger proportion of the total computational cost. We conjecture that the attractiveness of binary representations is tightly coupled with high-dimensional models where a single prediction requires many floating point operations, such as convolutional neural networks. Relatively compact latent factor models do not fall into this category.

\section{Alternative approaches}
Given the limited success of binary embedding, a more promising approach is offered by existing work that focuses on reducing the number of dot products which need to be performed in order to retrieve top recommendations. \citet{koenigstein2012efficient} introduces an exact branch-and-bound based on metric trees as well as an approximate algorithm that clusters users and serves recommendations computed for the cluster centers. \citet{shrivastava2014asymmetric} extend well-known \citep{indyk1998approximate} locality-sensitive hashing techniques to maximum inner product search (MIPS) through asymmetric transformations of the query and candidate vectors.

The method that is most closely related to the model we present in this paper is \citet{Shen_2015_ICCV}, who expand on MIPS by learning asymmetric binary hash functions. The hash functions are trained to minimize the L2 norm between the inner product matrix of the original input vectors and the inner product matrix of their binary representations.

\section{Conclusion}
While prediction latency is a pressing concern for many recommender systems, we find that the already compact latent factor models do not stand to benefit from using binary latent representations. The sharp drop in representational power exhibited by binary embeddings makes other approaches (such as exact \citep{koenigstein2012efficient} and approximate \citep{shrivastava2014asymmetric} maximum inner product search) more promising avenues when optimizing large-scale ranking.

\bibliography{bibliography}


\begin{thebibliography}{00}


\ifx \showCODEN    \undefined \def \showCODEN     #1{\unskip}     \fi
\ifx \showDOI      \undefined \def \showDOI       #1{#1}\fi
\ifx \showISBNx    \undefined \def \showISBNx     #1{\unskip}     \fi
\ifx \showISBNxiii \undefined \def \showISBNxiii  #1{\unskip}     \fi
\ifx \showISSN     \undefined \def \showISSN      #1{\unskip}     \fi
\ifx \showLCCN     \undefined \def \showLCCN      #1{\unskip}     \fi
\ifx \shownote     \undefined \def \shownote      #1{#1}          \fi
\ifx \showarticletitle \undefined \def \showarticletitle #1{#1}   \fi
\ifx \showURL      \undefined \def \showURL       {\relax}        \fi
\providecommand\bibfield[2]{#2}
\providecommand\bibinfo[2]{#2}
\providecommand\natexlab[1]{#1}
\providecommand\showeprint[2][]{arXiv:#2}

\bibitem[\protect\citeauthoryear{Amatriain}{Amatriain}{2013}]%
        {amatriain2013big}
\bibfield{author}{\bibinfo{person}{Xavier Amatriain}.}
  \bibinfo{year}{2013}\natexlab{}.
\newblock \showarticletitle{Big \& personal: data and models behind netflix
  recommendations}. In \bibinfo{booktitle}{{\em Proceedings of the 2nd
  International Workshop on Big Data, Streams and Heterogeneous Source Mining:
  Algorithms, Systems, Programming Models and Applications}}. ACM,
  \bibinfo{pages}{1--6}.
\newblock


\bibitem[\protect\citeauthoryear{Courbariaux, Hubara, Soudry, El-Yaniv, and
  Bengio}{Courbariaux et~al\mbox{.}}{2016}]%
        {courbariaux2016binarized}
\bibfield{author}{\bibinfo{person}{Matthieu Courbariaux}, \bibinfo{person}{Itay
  Hubara}, \bibinfo{person}{Daniel Soudry}, \bibinfo{person}{Ran El-Yaniv},
  {and} \bibinfo{person}{Yoshua Bengio}.} \bibinfo{year}{2016}\natexlab{}.
\newblock \showarticletitle{Binarized neural networks: Training deep neural
  networks with weights and activations constrained to+ 1 or-1}.
\newblock \bibinfo{journal}{{\em arXiv preprint arXiv:1602.02830\/}}
  (\bibinfo{year}{2016}).
\newblock


\bibitem[\protect\citeauthoryear{Covington, Adams, and Sargin}{Covington
  et~al\mbox{.}}{2016}]%
        {covington2016deep}
\bibfield{author}{\bibinfo{person}{Paul Covington}, \bibinfo{person}{Jay
  Adams}, {and} \bibinfo{person}{Emre Sargin}.}
  \bibinfo{year}{2016}\natexlab{}.
\newblock \showarticletitle{Deep neural networks for youtube recommendations}.
  In \bibinfo{booktitle}{{\em Proceedings of the 10th ACM Conference on
  Recommender Systems}}. ACM, \bibinfo{pages}{191--198}.
\newblock


\bibitem[\protect\citeauthoryear{Dror, Koenigstein, Koren, and Weimer}{Dror
  et~al\mbox{.}}{2011}]%
        {dror2011yahoo}
\bibfield{author}{\bibinfo{person}{Gideon Dror}, \bibinfo{person}{Noam
  Koenigstein}, \bibinfo{person}{Yehuda Koren}, {and} \bibinfo{person}{Markus
  Weimer}.} \bibinfo{year}{2011}\natexlab{}.
\newblock \showarticletitle{The yahoo! music dataset and kdd-cup'11}. In
  \bibinfo{booktitle}{{\em Proceedings of the 2011 International Conference on
  KDD Cup 2011-Volume 18}}. JMLR. org, \bibinfo{pages}{3--18}.
\newblock


\bibitem[\protect\citeauthoryear{Harper and Konstan}{Harper and
  Konstan}{2016}]%
        {harper2016movielens}
\bibfield{author}{\bibinfo{person}{F~Maxwell Harper} {and}
  \bibinfo{person}{Joseph~A Konstan}.} \bibinfo{year}{2016}\natexlab{}.
\newblock \showarticletitle{The movielens datasets: History and context}.
\newblock \bibinfo{journal}{{\em ACM Transactions on Interactive Intelligent
  Systems (TiiS)\/}} \bibinfo{volume}{5}, \bibinfo{number}{4}
  (\bibinfo{year}{2016}), \bibinfo{pages}{19}.
\newblock


\bibitem[\protect\citeauthoryear{Hidasi, Karatzoglou, Baltrunas, and
  Tikk}{Hidasi et~al\mbox{.}}{2015}]%
        {hidasi2015session}
\bibfield{author}{\bibinfo{person}{Bal{\'a}zs Hidasi},
  \bibinfo{person}{Alexandros Karatzoglou}, \bibinfo{person}{Linas Baltrunas},
  {and} \bibinfo{person}{Domonkos Tikk}.} \bibinfo{year}{2015}\natexlab{}.
\newblock \showarticletitle{Session-based recommendations with recurrent neural
  networks}.
\newblock \bibinfo{journal}{{\em arXiv preprint arXiv:1511.06939\/}}
  (\bibinfo{year}{2015}).
\newblock


\bibitem[\protect\citeauthoryear{Kingma and Ba}{Kingma and Ba}{2014}]%
        {kingma2014adam}
\bibfield{author}{\bibinfo{person}{Diederik Kingma} {and}
  \bibinfo{person}{Jimmy Ba}.} \bibinfo{year}{2014}\natexlab{}.
\newblock \showarticletitle{Adam: A method for stochastic optimization}.
\newblock \bibinfo{journal}{{\em arXiv preprint arXiv:1412.6980\/}}
  (\bibinfo{year}{2014}).
\newblock


\bibitem[\protect\citeauthoryear{Koenigstein, Ram, and Shavitt}{Koenigstein
  et~al\mbox{.}}{2012}]%
        {koenigstein2012efficient}
\bibfield{author}{\bibinfo{person}{Noam Koenigstein},
  \bibinfo{person}{Parikshit Ram}, {and} \bibinfo{person}{Yuval Shavitt}.}
  \bibinfo{year}{2012}\natexlab{}.
\newblock \showarticletitle{Efficient retrieval of recommendations in a matrix
  factorization framework}. In \bibinfo{booktitle}{{\em Proceedings of the 21st
  ACM international conference on Information and knowledge management}}. ACM,
  \bibinfo{pages}{535--544}.
\newblock


\bibitem[\protect\citeauthoryear{Liu, Rogers, Shiau, Kislyuk, Ma, Zhong, Liu,
  and Jing}{Liu et~al\mbox{.}}{2017}]%
        {liu2017related}
\bibfield{author}{\bibinfo{person}{David~C Liu}, \bibinfo{person}{Stephanie
  Rogers}, \bibinfo{person}{Raymond Shiau}, \bibinfo{person}{Dmitry Kislyuk},
  \bibinfo{person}{Kevin~C Ma}, \bibinfo{person}{Zhigang Zhong},
  \bibinfo{person}{Jenny Liu}, {and} \bibinfo{person}{Yushi Jing}.}
  \bibinfo{year}{2017}\natexlab{}.
\newblock \showarticletitle{Related Pins at Pinterest: The Evolution of a
  Real-World Recommender System}. In \bibinfo{booktitle}{{\em Proceedings of
  the 26th International Conference on World Wide Web Companion}}.
  International World Wide Web Conferences Steering Committee,
  \bibinfo{pages}{583--592}.
\newblock


\bibitem[\protect\citeauthoryear{Lynch, Aryafar, and Attenberg}{Lynch
  et~al\mbox{.}}{2015}]%
        {lynch2015images}
\bibfield{author}{\bibinfo{person}{Corey Lynch}, \bibinfo{person}{Kamelia
  Aryafar}, {and} \bibinfo{person}{Josh Attenberg}.}
  \bibinfo{year}{2015}\natexlab{}.
\newblock \showarticletitle{Images Don't Lie: Transferring Deep Visual Semantic
  Features to Large-Scale Multimodal Learning to Rank}.
\newblock \bibinfo{journal}{{\em arXiv preprint arXiv:1511.06746\/}}
  (\bibinfo{year}{2015}).
\newblock


\bibitem[\protect\citeauthoryear{Mu{\l}a, Kurz, and Lemire}{Mu{\l}a
  et~al\mbox{.}}{2016}]%
        {mula2016faster}
\bibfield{author}{\bibinfo{person}{Wojciech Mu{\l}a}, \bibinfo{person}{Nathan
  Kurz}, {and} \bibinfo{person}{Daniel Lemire}.}
  \bibinfo{year}{2016}\natexlab{}.
\newblock \showarticletitle{Faster Population Counts using AVX2 Instructions}.
\newblock
  \bibinfo{howpublished}{\url{https://github.com/kimwalisch/libpopcnt}}.
\newblock \bibinfo{journal}{{\em arXiv preprint arXiv:1611.07612\/}}
  (\bibinfo{year}{2016}).
\newblock


\bibitem[\protect\citeauthoryear{Rastegari, Ordonez, Redmon, and
  Farhadi}{Rastegari et~al\mbox{.}}{2016}]%
        {rastegari2016xnor}
\bibfield{author}{\bibinfo{person}{Mohammad Rastegari},
  \bibinfo{person}{Vicente Ordonez}, \bibinfo{person}{Joseph Redmon}, {and}
  \bibinfo{person}{Ali Farhadi}.} \bibinfo{year}{2016}\natexlab{}.
\newblock \showarticletitle{Xnor-net: Imagenet classification using binary
  convolutional neural networks}. In \bibinfo{booktitle}{{\em European
  Conference on Computer Vision}}. Springer, \bibinfo{pages}{525--542}.
\newblock


\bibitem[\protect\citeauthoryear{Rendle, Freudenthaler, Gantner, and
  Schmidt-Thieme}{Rendle et~al\mbox{.}}{2009}]%
        {rendle2009bpr}
\bibfield{author}{\bibinfo{person}{Steffen Rendle}, \bibinfo{person}{Christoph
  Freudenthaler}, \bibinfo{person}{Zeno Gantner}, {and} \bibinfo{person}{Lars
  Schmidt-Thieme}.} \bibinfo{year}{2009}\natexlab{}.
\newblock \showarticletitle{BPR: Bayesian personalized ranking from implicit
  feedback}. In \bibinfo{booktitle}{{\em Proceedings of the twenty-fifth
  conference on uncertainty in artificial intelligence}}. AUAI Press,
  \bibinfo{pages}{452--461}.
\newblock


\bibitem[\protect\citeauthoryear{Shrivastava and Li}{Shrivastava and
  Li}{2014}]%
        {shrivastava2014asymmetric}
\bibfield{author}{\bibinfo{person}{Anshumali Shrivastava} {and}
  \bibinfo{person}{Ping Li}.} \bibinfo{year}{2014}\natexlab{}.
\newblock \showarticletitle{Asymmetric LSH (ALSH) for sublinear time maximum
  inner product search (MIPS)}. In \bibinfo{booktitle}{{\em Advances in Neural
  Information Processing Systems}}. \bibinfo{pages}{2321--2329}.
\newblock


\bibitem[\protect\citeauthoryear{Weston, Bengio, and Usunier}{Weston
  et~al\mbox{.}}{2011}]%
        {weston2011wsabie}
\bibfield{author}{\bibinfo{person}{Jason Weston}, \bibinfo{person}{Samy
  Bengio}, {and} \bibinfo{person}{Nicolas Usunier}.}
  \bibinfo{year}{2011}\natexlab{}.
\newblock \showarticletitle{Wsabie: Scaling up to large vocabulary image
  annotation}. In \bibinfo{booktitle}{{\em IJCAI}}, Vol.~\bibinfo{volume}{11}.
  \bibinfo{pages}{2764--2770}.
\newblock


\end{thebibliography}
\bibliographystyle{ACM-Reference-Format}

\end{document}